# Optical and electronic properties in amorphous BaSnO$_3$ thin films

Jing Wang[1, 2], Bingcheng Luo[1, a]

[1] School of Physical Science and Technology, Northwestern Polytechnical University, Xi'an, Shaanxi, 710072, China

[2] Department of Physics, Weinan Normal University, Weinan, Shaanxi 714000, China

[a] Corresponding author

Address: Mailbox709, School of Physical Science and Technology, Northwestern Polytechnical University, Xi'an, Shaanxi, 710072, China

Tel: +86-029-88431670

Electronic mail: luobingcheng@nwpu.edu.cn

**Abstract**

Wide-bandgap perovskite stannates are of interest for the emergent all-oxide transparent electronic devices due to their unparalleled room temperature electron mobility. Considering the advantage of amorphous material in integrating with non-semiconductor platforms, we herein reported the optical and electronic properties in the prototypical stannate, amorphous barium stannate (BaSnO$_3$) thin films, which were deposited at room temperature and annealed at various temperatures. Despite remaining amorphous status, with increasing the annealing temperature, the defect level within amorphous BaSnO$_3$ thin films could be suppressed. Accordingly, the optical band gap, the leakage current, and the breakdown electric field could be improved from 2.81 eV, 3.94×10$^{-6}$ A cm$^{-2}$ at 400 kV/cm, 1.43 MV/cm for the as-prepared BaSnO$_3$ thin films to 2.90 eV, 6.17×10$^{-8}$ A cm$^{-2}$ at 400 kV/cm, 1.93 MV/cm for the 400 ℃-annealed BaSnO$_3$ thin films.

**Introduction**

Transparent amorphous oxides have attracted tremendous attention for applications in electronic and optoelectronic devices due to their advantages including the low processing temperature, the good uniform and the highly compatible with non-semiconductor platform, in comparison with crystalline counterparts. Amorphous indium-gallium-zinc-oxide (*a*-IGZO) is one of the most successful transparent amorphous oxides, which is widely used in commercially available panel displays [1, 2]. However, both indium and gallium in *a*-IGZO are rare and expensive, therefore, the development for resource-abundant transparent amorphous oxides is of interest.

Stannate oxides with earth-abundant elements have gained a good deal of attention for potential applications from gas sensors to transparent electronic devices [3-6]. Within this class of oxides, barium stannate ($BaSnO_3$, BSO), the archetypical cubic perovskite structure with a wide band gap (>3 eV), appears one of the most exciting objects [5]. Specifically, the highest room temperature (RT) electron mobility (~320 $cm^2V^{-1}s^{-1}$) among all the perovskite oxides and the exceptional oxygen stability (up to 530 °C) were observed when low-amount of lanthanum were doped into BSO single crystals [7-9]. These exciting properties are prompting the expansion of research enthusiasm on *n*-type BSO thin films. Typically, the enhanced RT device performances were realized when using La-doped BSO respectively as a channel layer and as an electron transport layer in BSO-based field-effect transistors and hybrid perovskite solar cells, both of which are in the limelight of perovskite-based device researches [10-13].

Past studies for BSO-based thin films focused almost exclusively on their crystalline forms [8-12, 14-24], which are processed under high temperature (> 600 °C) conditions and thus challenge the current semiconductor technology. Accordingly, an open question is whether amorphous BSO is a promising alternative for device applications, but amorphous BSO is less understood than the crystalline counterparts. Notably, it was quite recently shown that amorphous (Zn,Ba)SnO$_3$-based thin-film transistors exhibit the high field-effect mobility (>20 cm$^2$V$^{-1}$s$^{-1}$) and the good photo-stability [25]. However, the fundamental material properties of amorphous BSO thin films were not involved previously and needs to be investigated further. Additionally, the defects in BSO thin films could modify the electronic structure by the induced in-gap states and provide the shallow/deep trap centers to affect the optoelectronic properties [26]. Due to the natural presence of defects in amorphous BSO thin films, a better understanding of the charge carrier traps involving defects is required for further improving their performance. Accordingly, in this work, we attempt to understand the relationship between defect and properties by analyzing the optical and electrical characteristics of amorphous BSO thin films, which would provide opportunities for engineering the electronic properties and is of importance for stannate-based materials in next-generation electronic device applications.

**Experimental**

Amorphous BaSnO$_3$ (*a*-BSO) thin films with thickness of ~50 nm were deposited on double-side polished quartz and Pt-coated silicon (*i.e.*, (111) Pt/Ti/SiO$_2$/Si) substrates (MTI Corp.) at room temperature in one-shot by radio-frequency (RF) magnetron sputtering technique. The ceramic BaSnO$_3$ target was used and the detailed deposition conditions were described elsewhere [26, 27]. Briefly, the *a*-BSO thin films were fabricated under a power intensity of 2.5 W/cm$^2$, and a mixed gas atmosphere of argon (Ar) and oxygen (O$_2$) with a ratio of 4:1. After deposition, the as-prepared thin films were annealed at 200 ℃ and 400 ℃ in air for 20 min, respectively, both of which are lower than the crystallization temperature of BSO [28]. For brevity, the abbreviated symbol "T*x*" was used to represent different annealing samples, for example, T200 corresponding to BSO thin film annealed at 200 ℃. The *a*-BSO thin films deposited on quartz substrates were used for optical characterization, *i.e.*, transmittance and absorption spectra, which was performed by a U3010 spectrophotometer with an integrating sphere using quartz substrate as the reference. The *a*-BSO thin films deposited on Pt-coated silicon substrates with a Pt/*a*-BSO/Pt vertical structure were used for electrical measurement, which was carried out by a keithley 6517 meter. The size of top Pt square electrode is 0.2mm×0.2mm. The structure of *a*-BSO thin films was examined by grazing incidence X-ray diffraction (GIXRD) (PANalytical Empyrean). X-ray photoelectron spectroscopy (XPS) (ESCALAB 250) was used to probe the element valence states through narrow scan spectra of Ba3*d*, Sn3*d* and O1*s*, and the corresponding peak position was calibrated by the C1*s* line (284.6 eV).

**Results and discussion**

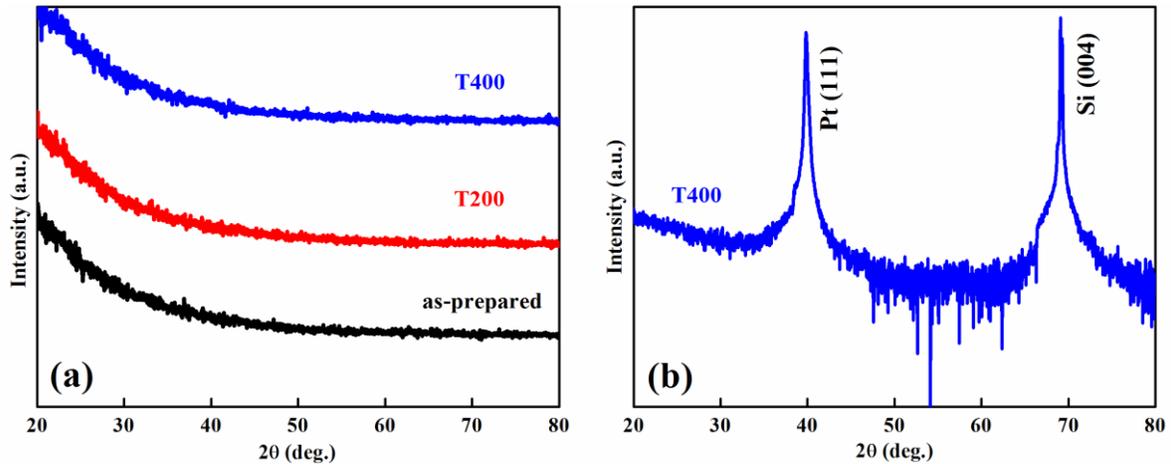

*Figure 1 XRD spectra for a-BSO thin films deposited on Pt-coated silicon substrates, (a) GIXRD patterns, (b) Normal θ-2θ XRD patterns for the T400 sample.*

Figure 1 (a) presents the GIXRD patterns of the BSO thin films deposited on Pt-coated silicon substrates. No obvious diffraction peaks are observed regardless of annealing temperature, indicating the amorphous feature as expected. This result is also supported by the normal θ–2θ

XRD patterns of BSO thin films, as shown in Figure 1 (b), which only shows the diffraction peaks referring to Pt (111) and Si (004).

Figure 2 (a) shows the XPS spectra collected in the Ba3$d$ regime. The double spin-orbital coupled spectral lines are evident for all the thin films, respectively corresponding to Ba3$d_{5/2}$ and Ba3$d_{3/2}$. The spin-orbital splitting energy is about 15.3 eV, which is consistent with the value expected for a Ba$^{2+}$ oxide [29]. Moreover, the peak position in Ba3$d$ spectrum is nearly identical for all the $a$-BSO thin films, indicating the similar chemical environment for Ba species. Additionally, one may note that the Ba3$d_{5/2}$ peak exhibits the asymmetry on the high binding energy side. The similar behavior was reported in BSO single crystal substrate [30], which is mainly attributed to the modified chemical environment at the topmost surface with respect to the bulk.

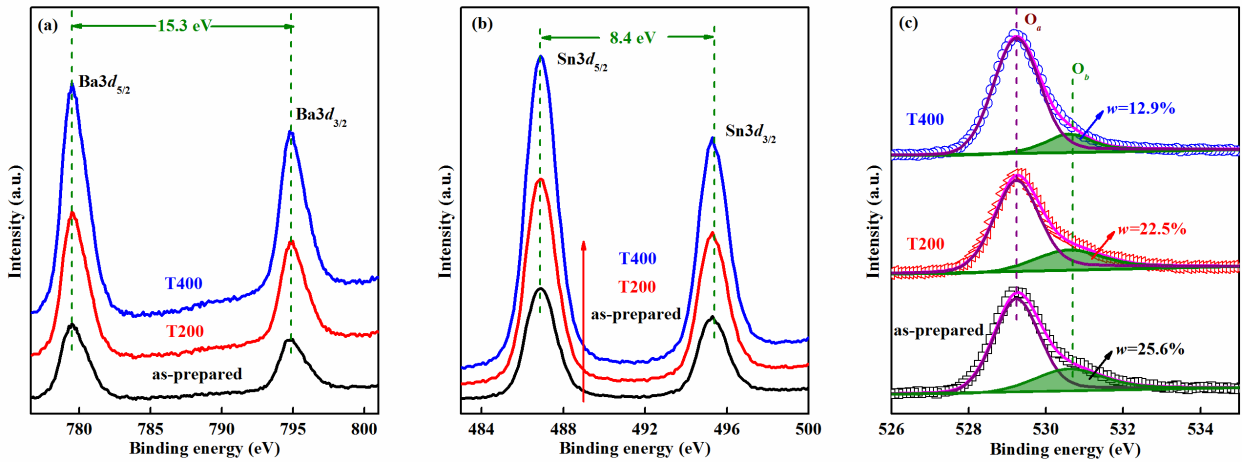

*Figure 2 XPS narrow-scan spectra for a-BSO thin films deposited on Pt-coated silicon substrates, (a) Ba3d, (b) Sn3d, (c) O1s. The solid lines in (c) represent the fitting results.*

The XPS signals collected in the Sn3$d$ regime, as shown in Figure 2 (b), exhibit two primary peaks for all the thin films, respectively corresponding to Sn3$d_{5/2}$ and Sn3$d_{3/2}$, due to the spin-orbital coupled effect. The Sn3$d$ spin-orbital splitting value of 8.4eV and the Sn3$d_{5/2}$ binding energy values of 486.8 eV are similar to Sn ions in the crystalline perovskite phases of barium stannate with a Sn$^{4+}$ state [31]. This indicates that the SnO$_6$ octahedron could be formed despite the absence of long-range order in the present case. Moreover, despite the various annealing temperatures, no obvious shift of binding energy of Sn3$d_{5/2}$ is observed, indicating the similar chemical environment for Sn$^{4+}$ ions. This is not similar to that reported in crystalline BSO thin films [26], probably due to the low annealing temperature in the present case.

The XPS O1$s$ spectra for $a$-BSO thin films are shown in Figure 2 (c), wherein every peak could be deconvoluted into two Gaussian–Lorentzian peaks. The low binding energy peak (O$_a$) at 529.2 eV is attributed to metal–oxide bonding in the perovskite phase, and the high binding energy peak (O$_b$) at 530.6 eV is usually supposed to be correlated with the chemisorbed oxygen, typical from a hydroxyl group [31]. For $a$-BSO thin films with an imperfectly microstructure, there is a large number of defects including dangling bonds and oxygen vacancies, which provide the absorption centers for the hydroxyl groups. Accordingly, the variation of O$_b$ peak would reflect the varying density of defects. As seen in Figure 2 (c), the ratio of O$_b$ peak area ($w$=O$_b$/(O$_a$+O$_b$)) decreases significantly with increasing the annealing temperature, meaning that there is a lower content of defects in these annealed samples relative to those in the as-prepared $a$-BSO thin films, despite remaining amorphous status.

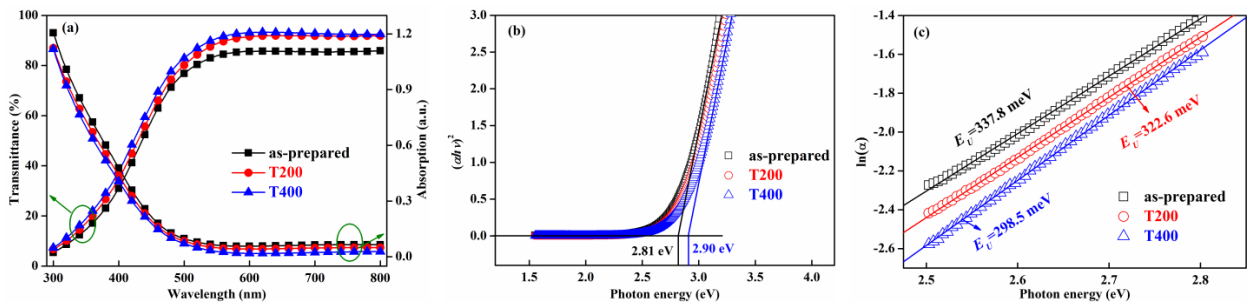

*Figure 3 Optical features for a-BSO thin films deposited on quartz substrates, (a) Transmittance and absorption spectra, (b) Tauc plots from absorption spectra, (c) The logarithm of absorption spectra as a function of photon energy below the band edge. The solid lines in (c) represent the fitting results.*

Figure 3 (a) displays the transmittance *(T)* and absorption *(α)* spectra for *a*-BSO thin films in the wavelength range of 300-800 nm. The average transmittance of all the *a*-BSO thin films is over 80% in the visible regime. In comparison to the as-prepared film, the annealed samples exhibit higher optical transparency in the visible regime. Defects, typically like oxygen vacancies in stannate-based oxides, are commonly inevitable, which could form subgap states near the conduction edge [26, 32-35]. Upon annealing, the density of defects is reduced, as above-mentioned in XPS analysis. This means that the subgap absorption originating from electron transitions from the valence band to these subband states in the annealed samples is suppressed, thereby leading to the increased transparency.

According to the absorption spectra in Figure 3 (a), one can estimate the optical band gap $E_g$ through extrapolation of the linear part to the photon energy axis in the Tauc plot, $(\alpha*E)^2$ *vs. E*, where *α* is the absorption coefficient, and *E* is the incident photon energy. As shown in Figure 3 (b), the $E_g$ values from 2.81 to 2.90 eV, along with the blue-shifted absorption edges in the annealed samples are observed. These $E_g$ values for *a*-BSO thin films are lower than those of crystalline BSO thin films [26]. Commonly, the presence of the band-tail states in amorphous status due to higher lattice disorder could lower the band-to-band transition energy, relative to the crystalline counterparts [36]. Upon annealing, the structural defects are reduced and better microstructure is obtained, so the width of the band-tail states shrinks and the optical band gap is widened. These band-tail states near the band edge can be evaluated by the Urbach relation, $\alpha \propto \exp(E/E_U)$, where $E_U$ is the Urbach energy. As shown in Figure 3 (c), the $E_U$ values are extracted to be about 337.8, 322.6, and 298.5 meV for the as-deposited, T200 and T400 samples, respectively. The Urbach energy decreases with increasing the annealing temperature, evidently reflecting that thermal annealing could improve the microstructure and reduce the tail-like absorption in the annealed samples.

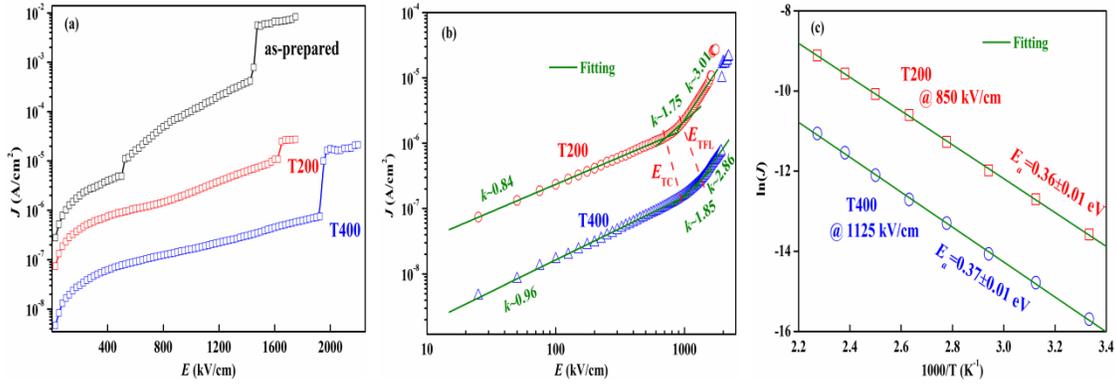

Figure 4 Leakage characteristics for a-BSO thin films deposited on Pt-coated silicon substrates, (a) current density J vs. electric field E, (b) J-E curves with a double-logarithmic scale, (c) Arrehenius plots in the Child's law region. The solid lines in (b) and (c) represent the fitting results.

Figure 4 (a) shows the leakage current density-electric field (*J-E*) characteristics for Pt/*a*-BSO/Pt devices. Evidently, the leakage current density is suppressed significantly with increasing the annealing temperature, typically from $3.94 \times 10^{-6}$ A cm$^{-2}$ in the as-prepared sample to $6.17 \times 10^{-8}$ A cm$^{-2}$ in the 400 °C-annealed sample at 400 kV/cm. The reduced density of defects, as demonstrated above, is the dominant cause of this behavior, because defects can provide the conduction path for field-assisted carrier transport through the trapping/detrapping processes [37]. Additionally, the current density increases suddenly at certain electric fields, for example, 1.43, 1.63, and 1.93 MV/cm for the as-deposited, T200 and T400 samples, respectively. This may be an implication of the soft breakdown phenomenon [38].

To better uncover the conduction mechanism, the same data in Figure 4 (a) are replotted with a double-logarithmic scale, as shown in Figure 4 (b). Here, only the T200 and T400 samples are considered due to the relatively complex soft breakdown behavior in the as-prepared sample. These plots demonstrate that the data match well with the space-charge-limited current (SCLC) model [39, 40]. Specifically, three linear regimes with different slopes ($k$) are observed before breakdown, respectively corresponding to Ohm's law regime, modified Child's law regime, and trap-filled-limited (TFL) regime [39, 40]. The transition electric fields ($E_{TC}$ and $E_{TFL}$) between the adjacent regimes are marked in Figure 4 (b). The presence of defects in these $a$-BSO thin films is demonstrated above, which commonly acts as carrier traps. In the low field regime ($E<E_{TC}$), the thermally generated free carriers dominate the leakage current, thereby the curves vary according to Ohmic-like ($J \propto E^m$, $m \sim 1$) conduction. With increasing electric fields ($E_{TC}<E<E_{TFL}$), more carriers are injected into the $a$-BSO thin films, and parts of the available trap states are filled, thereby the curves are consistent with the modified Child's law ($J \propto E^m$, $m>1$) conduction. When the field approaches $E_{TFL}$, all the trap states in $a$-BSO thin films would be gradually filled up by the injected carriers. Accordingly, the carriers injected later would move freely, thereby resulting in the increased slopes in the curves. However, when the electric field is over a certain value, the current density increases suddenly, indicating the formation of a conductive path probably due to the migration of oxygen vacancies under the electric field. Such behaviors are widely observed in filament-type oxide memristors containing oxygen vacancies [41], and further work needs to be done.

Note that all the trap states are completely filled up at $E_{TFL}$, therefore, the total density of trap states $N_t$ can be estimated according to the equation $N_t = 2\varepsilon_0\varepsilon_r E_{TFL}/(qd)$ [39], where $\varepsilon_0$ is the vacuum permittivity, $\varepsilon_r$ is the dielectric constant of BSO, $q$ is the elemental charge, and $d$ is the film thickness. The $N_t$ values are calculated to be $6.7 \times 10^{17}$ cm$^{-3}$ and $4.6 \times 10^{17}$ cm$^{-3}$ for T200 and T400 samples (here $\varepsilon_r=15$) [42], respectively, and this trend matches well with the qualitative analysis as mentioned above. Furthermore, the trap energy level ($E_t$) could be extracted through measuring the temperature-dependent current density in the Child's law regime. Typical Arrhenius plots of ln($J$) vs. 1000/$T$ are shown in Figure 4 (c). From the linear fitting, the slopes corresponding to ($E_t$-$E_c$)/1000$k_B$, ($E_c$: the minimum of the conduction band, $k_B$: the Boltzmann constant), are determined, and the activation energies ($E_a=E_c-E_t$) are calculated to be about 0.36±0.01 eV and 0.37±0.01 eV for T200 and T400 samples, respectively. Such $E_a$ values are similar to the reported deep energy levels resulting from oxygen vacancies [26, 32, 40]. Also, it is interesting to note that the $E_a$ values change little from T200 to T400 samples, in spite of different defect densities as demonstrated above. Accordingly, one can reasonably conclude that the similar $E_a$ values in two samples are mainly related to the electronic transition from the deep energy levels involving the ionized oxygen vacancies to conduction band.

**Conclusions**

In conclusion, amorphous BaSnO$_3$ (*a*-BSO) thin films with thickness of ~50 nm were fabricated at room temperature by RF magnetron sputtering technique, and the influence of post annealing temperature on their optical and electronic properties was investigated. It is found that the post annealing temperature significantly affects the properties of BSO thin films, in spite of remaining amorphous status. With increasing the annealing temperature, the defect level within the *a*-BSO thin films is suppressed, the optical band gap is widened, the leakage current is reduced, and the breakdown electric field is enhanced. Typically, the $a$-BSO thin film annealed at 400 °C has an optical band gap $E_g$ of ~2.90 eV, an Urbach energy $E_U$ of ~298.5 meV, a leakage current $J$ of $6.17 \times 10^{-8}$ A cm$^{-2}$ at 400 kV/cm, a breakdown electric field $E_b$ of 1.93 MV/cm, and a defect density $N_t$ of $4.6 \times 10^{17}$ cm$^{-3}$. Additionally, analysis of the leakage current behaviors suggests the dominance of space-charge-limited current mechanisms in the $a$-BSO thin films and yields a trap deep energy level of ~0.37 eV below conduction band, which is mainly ascribed to the presence of oxygen vacancies. These results reflect the importance of post annealing temperature in the modified properties of $a$-BSO thin films, which is valuable for amorphous stannate-based devices integrating on the non-semiconductor platforms.

## Acknowledgements

This work was supported by the Fundamental Research Funds for the Central Universities (No.310201911cx024).

# Figure captions

Figure 1 XRD spectra for *a*-BSO thin films deposited on Pt-coated silicon substrates, (a) GIXRD patterns, (b) normal $\theta$-$2\theta$ XRD patterns for the T400 sample.

Figure 2 XPS narrow-scan spectra for *a*-BSO thin films deposited on Pt-coated silicon substrates, (a) Ba3*d*, (b) Sn3*d*, (c) O1*s*. The solid lines in (c) represent the fitting results.

Figure 3 Optical features for *a*-BSO thin films deposited on quartz substrates, (a) Transmittance and absorption spectra, (b) Tauc plots from absorption spectra, (c) The logarithm of absorption spectra as a function of photon energy below the band edge. The solid lines in (c) represent the fitting results.

Figure 4 Leakage characteristics for *a*-BSO thin films deposited on Pt-coated silicon substrates, (a) Current density *J* *vs.* electric field *E*, (b) *J-E* curves with a double-logarithmic scale, (c) Arrehenius plots in the Child's law region. The solid lines in (b) and (c) represent the fitting results.